\documentclass[journal]{IEEEtran}

\usepackage{booktabs}
\usepackage{cite}
\usepackage{bbm}
\usepackage{amsmath,amssymb,amsfonts}
\usepackage{algorithmic}
\usepackage{graphicx}
\usepackage{textcomp}
\usepackage{xcolor}
\usepackage{amsmath}
\usepackage{diagbox}
\usepackage[misc]{ifsym}
\usepackage{mathrsfs}
\usepackage{float}
\usepackage{multirow}
\usepackage{subfigure}

\makeatletter\renewcommand{\maketag@@@}[1]{\hbox{\m@th\normalsize\normalfont#1}}%
\makeatother

\begin{document}
\title{Leveraging Multiple Legacy Wi-Fi Links for Human Behavior Sensing}

\author{Lingchao Guo,
Zhaoming Lu\textsuperscript{\dag}, \emph{Member, IEEE,}
Xiangming Wen, \emph{Senior Member, IEEE,}\\
Liming Wang,
David Gesbert, \emph{Fellow, IEEE,}
and Zijun Han
\thanks{Lingchao Guo, Zhaoming Lu, Xiangming Wen, Liming Wang, and Zijun Han are with Beijing Key Laboratory of Network System Architecture and Convergence, Beijing Laboratory of Advanced Information Networks, and the School of Information and Communication Engineering, Beijing University of Posts and Telecommunications, Beijing, 100876, China. E-mail:\{rita\_guo, lzy0372, xiangmw, wangliming, zijunhan\}@bupt.edu.cn.}
\thanks{David Gesbert is with the Communications Systems Department, EURECOM, Sophia Antipolis, France. E-mail: David.Gesbert@eurecom.fr.}
\thanks{\textsuperscript{\dag}Corresponding author.}
\thanks{A part of this work was accepted for publication in 2021 IEEE Wireless Communications and Networking Conference (WCNC), 2021 \cite{shen2021wiagent}.}
\thanks{The work of David Gesbert was partially funded via the HUAWEI France supported Chair on Future Wireless Networks at EURECOM.}
}

\maketitle

\begin{abstract}

Taking advantage of the rich information provided by Wi-Fi measurement setups, Wi-Fi-based human behavior sensing leveraging Channel State Information (CSI) measurements has received a lot of research attention in recent years. The CSI-based human sensing algorithms typically either rely on an explicit channel propagation model or, more recently, adopt machine learning so as to robustify feature extraction. In most related work, the considered CSI is extracted from a single dedicated Access Point (AP) communication setup. In this paper, we consider a more realistic setting where a legacy network of multiple APs is already deployed for communications purposes and leveraged for sensing benefits using machine learning. The use of legacy network presents challenges and opportunities as many Wi-Fi links can present with richer yet unequally useful data sets. In order to break the curse of dimensionality associated with training over a too large dimensional CSI, we propose a link selection mechanism based on Reinforcement Learning (RL) which allows for dimension reduction while preserving the data that is most relevant for human behavior sensing.
The method is based on a sequential state decision-making process in which the CSI is modeled as a part of the state. From actual experiment results, our method is shown to perform better than state-of-the-art approaches in a scenario with multiple available Wi-Fi links.


\end{abstract}


\IEEEpeerreviewmaketitle

\section{Introduction}

 \IEEEPARstart{W}{ith} the ubiquitous deployment of Wi-Fi infrastructure and aided by the Channel State Information (CSI) made accessible in commodity Wi-Fi chipsets \cite{halperin2011tool}, Wi-Fi-based human behavior sensing has become an active research area in both industry and academia \cite{yousefi2017a, wang2018wi, guo2017behavior, guo2018behavior, chen2017human}. Applications range from human detection \cite{zhu2017r, qian2018enabling, gu2017mosense}, fall detection \cite{palipana2018falldefi, cheng2019deep}, human authentication \cite{kong2020continuous, liu2018authenticating, cao2021a, korany2021multiple}, gesture recognition \cite{ma2018signfi,  yang2019learning, venkatnarayan2020WiFi, venkatnarayan2021wifi}, daily activity recognition \cite{wang2017device, jiang2018towards, gao2017csi, xiao2018seare, wang2019on, zhang2019WiFimap, feng2019wi, zhang2021data} to human imaging \cite{guo2020from, jiang2020towards, wang2021from}. Further, in a dynamic setting, it is in principle possible to monitor changes in human behaviors by leveraging the temporal dimension of Wi-Fi CSI. Wi-Fi CSI-based human behavior sensing offers many advantages, including being a low-cost, non-invasive, and highly practical solution for smart homes, elderly monitoring, intrusion detection, among others.


CSI-based human behavior sensing is rooted in the principle of multipath imaging, as human bodies can absorb or bounce radio waves thereby modifying our perception of multipath. CSI hence captures the various reflection, diffraction and scattering effects, which can in turn be traced back to a certain body location and pose.

\begin{figure}[t]
  \centering{\includegraphics[width=0.99\linewidth]{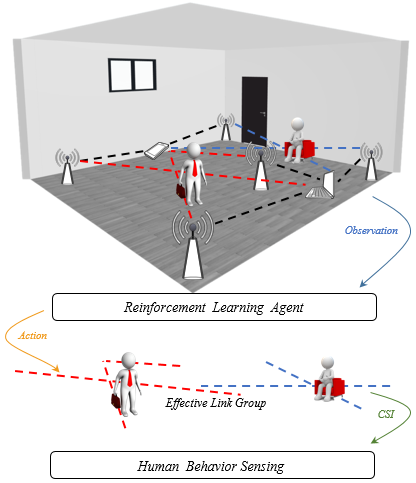}}
  \caption{An illustration of the legacy Wi-Fi network-based sensing framework. At each time step, a reinforcement learning agent proposes a selection for the most suitable subset of Wi-Fi links towards the sensing goals. At convergence, the CSI of the selected links is processed for human behavior sensing.}
  \label{first}
\end{figure}

Previous Wi-Fi CSI-based human behavior sensing systems can be roughly divided into two categories according to the types of the algorithms used: modeling-based systems and learning-based systems.
To be specific, modeling-based systems capture human movement through Wi-Fi CSI by building an explicit propagation model \cite{wang2015understanding, zhang2019towards, wu2020gaitway, qian2017inferring, pu2013whole, shen2019wirim, arshad2017wi} while learning-based systems distinguish human behaviors through features extracted from Wi-Fi CSI based on machine learning \cite{jiang2018towards, wang2018device, qian2018enabling, wang2017wifall, fei2020multi, li2018learning, wang2014e, huang2020towards, li2016WiFinger, ali2017recognizing, guo2020wireader, guo2020from, jiang2020towards, wang2021from}.

In most cases so far, these systems have been investigated in simple Wi-Fi environments, typically  consisting of a single dedicated Access Point (AP) and one or several Wi-Fi receivers (Rxs). Furthermore, the Wi-Fi devices are often deployed at locations that are intuitively perceived as potentially favorable for human sensing purposes. The most common one is that an AP and two Rxs forming two orthogonal segments in space \cite{qian2017inferring, guo2020from, wang2021from}, as a pair of transceiver can only sense the motion component perpendicular to the Line of Sight (LoS) path between them according to \cite{8067692}.

The physical modeling approach is attractive in settings where one can get away with a small number of parameters, for instance simple propagation (single bounce) environments with one (or a small number of) Wi-Fi device(s). However, since we are likely to see a larger number of APs and more complex propagation settings, a physical modeling approach is unpractical, making a learning approach the preferred option \cite{li2018learning}. Unfortunately, the learning task in the case of a possibly large network of Wi-Fi devices is made much more challenging for the reasons below.

One reason is the over-dimensioned nature of CSI data extracted from multiple access points, leading to increased training time and overall system complexity.
Furthermore, realistic Wi-Fi deployments mean that APs are located non-optimally from a human sensing point of view, resulting in certain links contributing mostly noise rather than meaningful data for the sensing and classification tasks at hand, in comparison to other links.



In this paper we ask the following question: Is there a way to leverage the rich CSI data arising from a legacy Wi-Fi network with multiple devices for human behavior sensing while mitigating the curse of dimensionality, i.e., keeping training time reasonable and pruning out noise-dominated data?

To answer this question, this paper proposes an approach based on the dynamic selection of Wi-Fi CSI data by a Reinforcement Learning (RL) agent. Intuitively, the principle is to decide and select in real-time the small subset of radio links that are the most relevant to human behavior sensing. Clearly, the difficulty of this task is the definition of a proper ``relevance" metric.

To achieve this goal, we construct a link selection algorithm that will help describe the most representative set of links for human behavior sensing. We formulate this link selection as a sequential decision making process, which is naturally modeled by RL \cite{sutton1988reinforcement}. The workflow of our method is illustrated in Figure \ref{first}.

At each time step, the RL agent receives the environment’s state, which relates to the human’s dynamic behavior. Relying on the received state, the agent chooses an action to the more suitable subset of radio links until convergence. The chosen action depends on the policy network, which is a probability distribution function that maps the current state to the action space. After a time step, we get a sensing reward as a result of the action, and the system transits to a new state. A carefully designed reward function is proposed according to sensing goals, and our objective is to prioritize the data from the set of links that maximize the total sum of the rewards.

Remarkably, we also develop a proof-of-concept prototype and apply the proposed link selection framework to recognize 5 activities at 16 different locations, with four radio links (corresponding to two Wi-Fi APs and two Wi-Fi Rxs at non-optimized locations). The prototype allows to test our algorithms, and compare them with relevant methods selected from the state-of-the-art (see section II for a description of such methods), hence showing the benefits of our approach.

The main contributions of this paper are summarized as follows.
	\begin{itemize}
		\item We propose a novel link selection mechanism for improving the performance of machine learning-based human behavior sensing in a scenario with multiple legacy Wi-Fi links. We intuitively formulate the link selection process as a sequential decision making process.
		\item RL is adopted to solve the formulated sequential decision problem. A novel framework that takes both context information and historical states into consideration for decision making is designed.
		\item For better decision making and human behavior sensing, a reward function is designed, together with an overall loss function based on a useful analogy with the problem of frame selection in the independent context of video processing \cite{wu2019multi}.
		\item We measure the performance of the RL-based framework with extensive actual experiments. We also compare our method with start-of-the-art methods. Experimental results attest the effectiveness of our framework.
	\end{itemize}

The rest of the paper is organized as follows. Related work is discussed in Section II. And the preliminaries and methodology are elaborated in Section III and Section IV, respectively. The implementation details of the experiments are presented in Section V. The performance evaluation is shown in Section VI, followed by some discussions and future work in Section VII.

	\begin{table*}[htbp]
	  \fontsize{8}{8}\selectfont
	  \setlength{\tabcolsep}{6pt}
	  \centering
	  \caption{Some existing representative Wi-Fi-based human behavior sensing works}
	    \begin{tabular}{ccccccc}
	    \toprule
	    \textbf{Category} & \textbf{Reference} & \textbf{Number of Txs} & \textbf{Number of Rxs} & \textbf{ Links} & \textbf{Learning-Assisted?} & \textbf{Applications} \\
	    \midrule
	    \multirow{6}[2]{*}{\textbf{Physical Modeling-Based}} & CARM \cite{wang2015understanding} & 1     & 1     & -- & \checkmark     & Activity Recognition \\
	          & \cite{zhang2019towards} & 1     & 1     & -- & Both  & Activity Recognition \\
	          & GAITWAY \cite{wu2020gaitway} & 1     & 1     & -- & \checkmark & Gait Recognition \\
	          & WiDance \cite{qian2017inferring} & 1     & 2     & Orthogonal & \checkmark     & Activity Recognition \\
	          & WiSee \cite{pu2013whole} & 1     & 1     & -- & \checkmark     & Gesture Recognition \\
	          & WiRIM \cite{shen2019wirim} & 1     & 1     & -- & \checkmark     & Activity Recognition \\
	          & Wi-Chase \cite{arshad2017wi} & 1     & 1     & -- & \checkmark     & Activity Recognition \\
	    \midrule
	    \multirow{8}[2]{*}{\textbf{Machine Learning-Based}} & PADS \cite{qian2018enabling} & 1     & 1     & -- & -- & Human Detection \\
	          & Wi-Fall \cite{wang2017wifall} & 2     & 2     & Two Links & -- & Fall Detection \\
	          & MAR \cite{fei2020multi} & 1     & 1     & -- & -- & Gait Recognition\\
	          & \cite{li2018learning} & 9     & 3     & Random & -- & Activity Recognition \\
	          & E-eyes \cite{wang2014e} & 1     & 3     & Random & -- & Activity Recognition \\
	          & WiAnti \cite{huang2020towards} & 1     & 1     & -- & -- & Activity Recognition \\
	          & WiFinger \cite{li2016WiFinger} & 1     & 1     & -- & -- & Gesture Recognition \\
	          & WiKey \cite{ali2017recognizing} & 1     & 1     & -- & -- & Keystroke Recognition \\
	          & WiReader \cite{guo2020wireader} & 1     & 1     & -- & -- & Handwriting Recognition \\
	          & Wi-Pose \cite{guo2020from} & 1     & 2     & Orthogonal & -- & Human Imaging \\
	          & WiPose \cite{jiang2020towards} & 1     & 3     & Random & -- & Human Imaging \\
						& Wi-Mose \cite{wang2021from} & 1     & 2     & Orthogonal & -- & Human Imaging \\
	    \bottomrule
	    \end{tabular}%
	  \label{tab:addlabel}%
	\end{table*}

\section{Related Work}
 In this section, we discuss some existing representative works from the two categories of Wi-Fi based sensing methodologies so far reported in the literature, namely physical modeling-based and machine learning-based.
\subsection{Physical Modeling-Based Systems}
 Wang et al. \cite{wang2015understanding} propose the CSI-Speed Model, which quantifies the correlation between CSI amplitude dynamics and the speed of path length change of the reflected paths caused by human movement. Based on the CSI-speed model and Hidden Markov Model (HMM), they build an activity recognition system called CARM which can differentiate and recognize eight predefined activities. Zhang et al. \cite{zhang2018from} \cite{niu2018a} develop the one-side and two-sides Fresnel diffraction model, which characterizes the relationship between the geometrical position of the sensing target and the induced CSI amplitude variations caused by the motion of the target. They demonstrate that the Fresnel diffraction model is effective and robust in recognizing exercises and daily activities \cite{zhang2019towards}. Wu et al. \cite{wu2020gaitway} present a scattering model, which treats environmental objects as multiple scatters and reveals that the CSI statistically embodies the target’s moving speed when accounting for a number of scattering multipaths. Upon the scattering model, they develop GAITWAY, which can monitor and recognize gait speed through the walls. Doppler phase shift \cite{qian2017inferring} \cite{pu2013whole}, Angle of Arrival (AoA) \cite{shen2019wirim} and Time of Flight (ToF) \cite{arshad2017wi} are also popular models for CSI-based human behavior sensing. However, due to the complexity of human behavior, these models still need machine learning methods to distinguish reliably between various activity classes.

 Specific information about the number and placement of transceivers of these systems are shown in Table \ref{tab:addlabel}, which indicates the above models are all built in simple Wi-Fi environments. At the same time, in order to realize the recognition of activities or gestures, they often need to do classification with the help of machine learning methods.

 \subsection{Machine Learning-Based Systems}
 There exists a significant body of work related to the the use of machine learning for Wi-Fi-aided human behavior classification. For instance, PADS \cite{qian2018enabling} uses One-Class Support Vector Machine (SVM) for human detection. Wi-Fall \cite{wang2017wifall} utilizes the k-Nearest Neighbors algorithm (kNN) and One-Class SVM for fall detection. MAR \cite{fei2020multi} tries to identify multiple-variations of body parts and applies CANDECOMP/ PARAFAC (CP) and Dynamic Time Wrapping (DTW) to recognize activities. \cite{li2018learning} designs a Convolutional Neural Network (CNN) model for activity recognition. E-eyes \cite{wang2014e} exploits Multi-Dimensional DTW and Pattern Matching to recognize household activities such as washing dishes and taking a shower. WiAnti \cite{huang2020towards} is a Wi-Fi-based human activity recognition system that is robust to Co-Channel Interference (CCI) and it evaluates the performance with six different classifiers. WiFinger \cite{li2016WiFinger} also utilizes Multi-Dimensional DTW and Pattern Matching and further designs a series of signal processing techniques to recognize finger gestures. WiKey \cite{ali2017recognizing} proposes to use kNN and DTW for keystroke detection and recognition. WiReader \cite{guo2020wireader} generates a energy feature matrix an combines with Long Short-Term Memory (LSTM) to realize the recognition of different handwriting actions. Wi-Pose \cite{guo2020from}, WiPose \cite{jiang2020towards}, and Wi-Mose \cite{wang2021from} exploit Cross-Modal networks to achieve 2D or 3D human pose estimation.

 Specific information about the number and placement of transceivers of these systems are also shown in Table \ref{tab:addlabel}. Most of the current machine learning-based works focus on simple Wi-Fi environments with typically just one Wi-Fi AP. One exception is \cite{li2018learning}, which suggests that machine learning-based approaches have the potential to sense human behaviors in multi-AP environments albeit at a high complexity cost, since \cite{li2018learning} does not prune out the links that bear little relevance to the actual classification task at hand.

 In contrast, in this work, we focus on possibly complex Wi-Fi deployments with multiple APs and highlight the importance of reducing the data dimensionality by selecting the subset of radio links that are most informative about the classification task. This problem is challenging as the Wi-Fi deployment is not optimized for this problem. We show however how this can be recast into a novel data-driven decision making framework exploiting RL.

 \begin{figure*}[!htpb]
	 \centering{\includegraphics[width=1\linewidth]{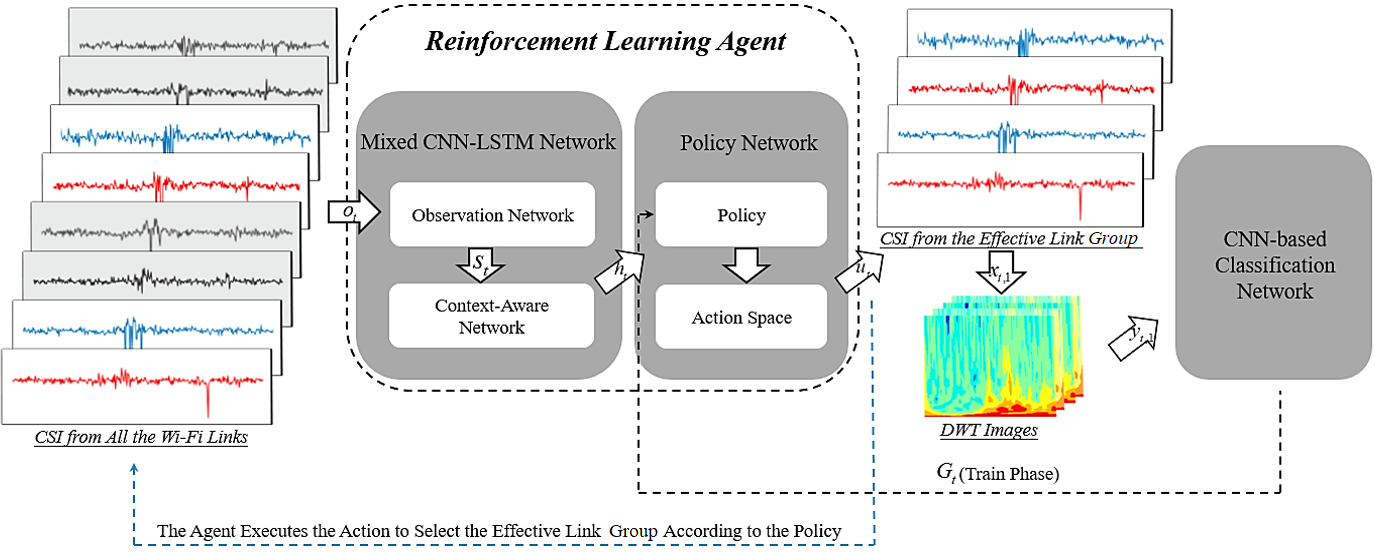}}
	 \caption{The proposed RL framework is consist of a mixed CNN-LSTM network for capturing environment state, a policy network for estimating probabilistic distribution over action space and a classification network for human behavior prediction.}
	 \label{frame}
 \end{figure*}
\section{preliminaries}
	\subsection{Principle of CSI-based Human Sensing}
  CSI characterizes how Wi-Fi signals propagate from the transmitter (Tx) to the Rx at different subcarriers. The basic form of CSI can be expressed as:
  \begin{equation}H_{\varsigma }=|H_{\varsigma }| e^{j \angle H_{\varsigma }}\end{equation}
  where amplitude $|H_{\varsigma }|$ and phase $\angle H_{\varsigma }$ at the ${\varsigma _{th}}$ subcarrier are impacted by the multipaths in the environment resulting in amplitude attenuation and phase shift. The presence of human or tiny body movement will affect the multipaths thus leading some changes to CSI. Assisted by physical modeling or machine learning algorithms, we can sense human behaviors by exploring the changes of CSI caused by humans. In this paper, we are considering the particular problem of classifying human behavior as one of a finite number of human activity classes (walk, run, stand, sit, and bend).

  The Linux CSI tool \cite{halperin2011tool} provides CSI for 30 out of 56 subcarriers for each antenna. Specifically, $N_{Tx} \times N_{Rx} \times 30$ CSI values are available in a sample, where $N_{Tx}$ and $N_{Rx}$ are the number of antennas in Tx and Rx respectively.
	However, given a complex Wi-Fi environment with $M$ Txs and $Q$ Rxs, there will be $M \times Q \times N_{Tx} \times N_{Rx} \times 30$ CSI values at each time, which will increase the difficulty of data processing and network complexity. Thus finding an effective way to extract the most representative features for human behavior sensing from such a huge amount of data is the key objective in this work.

	\subsection{Effective Link Selection}
A link refers to the combination of Wi-Fi multipath signals which correspond to a particular pair of transceivers. Given a complex Wi-Fi environment with $M$ Txs and $Q$ Rxs, as previously mentioned, there will be
$M \times Q$ links in the environment, leading to $M \times Q \times N_{Tx} \times N_{Rx} \times 30$ CSI values at each time sample. However, the intuition we put forward in this paper is that in a typical arbitrary Wi-Fi deployment scenario, the subset of links that are significantly affected by a single human behavior is limited. Dynamically picking out the links that are the most relevant to human behavior can significantly reduce the dimensionality of data processing at each time, hence improving classification performance for a given system complexity and training overhead.

Therefore, we firstly define the concept of effective link group to describe the most relevant set of links for human behavior sensing here.

\emph{\textbf{Notation:}} Given the set of APs \(\Phi {\rm{ = }}\left\{ {{\varphi _1},{\varphi _2},...,{\varphi _M}} \right\}\) and the set of Rxs
\(\Upsilon {\rm{ = }}\left\{ {{\upsilon _1},{\upsilon _2}, \ldots ,{\upsilon _Q}} \right\}\), the effective link group corresponding to a certain behavior $b$ performed at a certain location $l$  at the \({t_{th}}\) time step is expressed as
\({\Gamma _{lb}}\left( t \right) \subset \left\{ {\left( {{\varphi _m},{\upsilon _q}} \right)\left| {m = 1,2, \ldots ,M;q = 1,2, \ldots ,Q} \right.} \right\}\). Note that \({({\varphi _m},{\upsilon _q})}\) refers to the link between AP \({{\varphi _m}}\) and Rx \({{\upsilon _q}}\).

We now proceed to construct a RL agent, capable of identifying the most effective link group through a link selection algorithm.



\section{Methodology}
\subsection{Problem Formulation}
We propose an iterative algorithm for identifying the effective link group $\Gamma_{lb}\left( t \right)$.
Hence the agent initially considers all Wi-Fi links and iteratively selects individual links based on a reward maximization policy.


The process of link selection can be formalized as a Markov Decision Process (MDP), whereby a sequential decision-making problem is solved at each iteration. As is known, a generic MDP problem can be naturally solved by a RL agent which is further detailed below.

\subsection{Data Preprocessing}
Prior to link selection and to reduce the computation burden of the model and improve the accuracy, we preprocess the CSI of all the links, i.e., $M \times Q$ links, as follows.

Because a time series of CSI phase differences and amplitudes can detect the minute changes of the environment that alter signal propagation, we can use these measurements to capture relevant human behaviors \cite{shen2019wirim}.
Here, we first calculate $\Delta \angle {H_{\xi ,\varsigma  }}$, the phase differences between adjacent antennas. Then we apply the Principal Component Analysis (PCA) algorithm to the CSI streams $\left[|H_{\xi ,\varsigma }|, \Delta \angle H_{\xi ,\varsigma }\right]$ for each link $\xi$ ($\xi  = 1,2, \ldots ,M \times Q$), as PCA is widely used for helping fully extract the behavior-related features while greatly reducing the data dimensions and removing unrelated information \cite{guo2019wiroi}.
We use the second principal component for further processing since it clearly
captures human behaviors \cite{wang2015understanding}.
Let the two CSI streams of $\xi_{th}$ link be $x_{\xi ,t}=[x_{\xi ,t}^{1}, x_{\xi ,t}^{2}]$, where $x_{\xi, t}^{1}$ and $x_{\xi, t}^{2}$ is the second principal component of the phase difference $\Delta \angle H_{\xi ,\varsigma }$ and amplitude $|H_{\xi ,\varsigma }|$ at the time step $t_{th}$, respectively.

Then we choose the zero-mean normalization method to normalize the two CSI streams $x_{\xi, t}^{ i}$ with zero mean and unit variance to reduce the calculation cost and improve the classification performance, and then obtain $o_{\xi, t}^{i}$:
\begin{equation}o_{\xi, t}^{i}=\frac{x_{\xi, t}^{i}-\bar{x}_{\xi, t}^{i}}{\sigma_{\xi, t}^{ i}},\end{equation}
for $i=1,2$ and $\xi  = 1,2, \ldots ,M \times Q$,
where $\bar{x}_{\xi, t}^{i}$ and $\sigma_{\xi, t}^{i}$ are the mean and standard deviation of the $x_{\xi, t}^{i}$, respectively.

\subsection{Framework Design}
Figure \ref{frame} summarizes the overall framework, which consists of the following modules: an agent which decides the link group based on its context information together with a CNN-based classification network which is used to recognize the human behavior, but also to compute the RL reward.
To be specific, the agent exploits two networks: a mixed CNN-LSTM network which encodes the context information and historical states and also a policy network which generates a proper action from a predefined action space at each step.

In this section, we represent the framework by a tuple $\langle S, U, \mathcal{T}, G, \pi\rangle$, referring to states, actions, transition, reward, and policy, and we introduce the details for each of them one by one, as follows.

\subsubsection{State S}

	The RL agent first needs to encode the explored environment into a feature vector reflecting the context information.
	Therefore, we adopt an observation network $f_{o}$ and a context-aware network $f_{h}$ as the mixed CNN-LSTM network, parameterized by $\theta_{o}$ and $\theta_{h}$ respectively.

	The observation network $f_{o}$, composed of three convolution layers, is designed at the beginning of the mixed CNN-LSTM network to effectively reduce the dimension of the data and significantly simplify the training procedure. 
	At the time step $t_{th}$, the agent observes a state:
 \begin{equation}
 	s_{t}=f_{o}\left(o_{\xi, t}^{i}; \theta_{o}\right)=W_{o} x_{\xi, t}^{i}+b_{o},
\end{equation}
	for $i=1,2$ and $\xi  = 1,2, \ldots ,M \times Q$, where $f_{o}$ is a linear weighted sum function, weights $W_{o}$ and biases $b_{o}$ constitute the parameter $\theta_{o}$.

  Furthermore, on the one hand, human behavior sensing should be sensitively context-aware in our situation. On the other hand, multi-LSTM models are capable of learning and recognizing hidden patterns.
	Therefore, we design a context-aware network $f_{h}$, using three LSTM layers upon the observation network, to be aware of the context information and recognize the hidden patterns.
	The agent observes both a state $s_{t}$ and its previous hidden states $h_{t-1}$ as inputs of the context-aware network, then produces its current hidden states $h_{t}$:
\begin{equation}h_{t}=f_{h}\left(h_{t-1}, s_{t} ; \theta_{h}\right).\end{equation}

	\subsubsection{Action U}
	Action $U$ is the action space covers all the combination of Wi-Fi links, e.g., all possible link groups formed, and represents a set of discrete actions that our agent can take. At time step $t_{th}$, $u_{t} \in U$ decides a Wi-Fi link group from the set of possible actions according to the state $s_{t}$.

  \subsubsection{Transition $\mathcal{T}$}
	In response to the selected action $u_{t}$, function $\mathcal{T}: S \times U \rightarrow S$ maps a state ${s}_{t}$ into a new state ${s}_{t+1}$:
  \begin{equation}s_{t+1}=\mathcal{T}\left(s_{t}, u_{t}\right)\end{equation}

  \subsubsection{Policy $\pi$}
	Then the agent feeds the feature vector $h_{t}$ into the following policy network to generate a proper action from the action space. This action adjusts the link group that the agent decides at the next time.

	The policy network is composed of a fully connected layer parameterized by $\theta_{u}$ and uses a sigmoid activation function to output the probability value of each action.

	Clearly, the generation of an action indicates that the agent decides the link group corresponding to the action.
	The policy is as follows:
\begin{equation}\pi\left(u_{t} | h_{t} ; \theta_{u}\right)=sigmoid\left({W_{u}}h_{t}+{b_{u}}\right),u_{t} \in U,\end{equation}
	where $\pi\left(u_{t} | h_{t} ; \theta_{u}\right)$ is the action probability distribution, $sigmoid\left(\cdot \right)$ is a sigmoid function, ${W_{u}}$ and ${b_{u}}$ are the weights and biases, which together constitute the parameter $\theta_{u}$.
	The action with the highest probability, i.e., $u_{t}^{*}=argmax_{u_{t}}\pi\left(u_{t} | h_{t} ; \theta_{u}\right)$ is estimated as the proper action.

  \subsubsection{Reward G}
	When the link groups decided by agent keep unchanged after performing a series of actions, we consider the link group as the effective link group $\Gamma_{lb}\left( t \right)$ at this time. A classification network will emit the prediction based on the $\Gamma_{lb}\left( t \right)$.

The CNN-based classification network $f_{p}$ is parameterized by $\theta_{p}$.
Supposing there are a total of $\Omega$ links in $\Gamma_{lb}\left( t \right)$, for each link $\omega$
($\omega = 1,2,...,\Omega$), we firstly obtain the two CSI streams $x_{\omega, t}^{i}$ $(i=1,2)$ as mentioned before. Then we use Discrete Wavelet Transform (DWT) \cite{guo2020from} to extract the temporal-frequency features $y_{\omega, t}^{i}$ $(i=1,2)$ from CSI phase differences and amplitudes.
Based on this, we extract the DWT spectrum images at the ${t_{th}}$ time step from the link ${\omega}$, defined as:
\begin{equation}y_{\omega, t}^{i}=\textit{DWT}\left(x_{\omega, t}^{i}\right),\end{equation}
for  $i=1,2$ and $\omega  = 1,2, \ldots ,\Omega $.

	As the input data of the CNN-based classification network, the DWT images are then processed in convolution layers, dropout layers, maximum pooling layers and fully connected layers.
  The detailed architecture of the CNN-based classification network is presented in Section V. Note that we deploy the Rectified Linear Unit (ReLU) function in each convolution layer. The last fully connected layer uses a softmax function to output the probability value of each class.

	The predicted distribution according to link $\omega$ is:
  \begin{equation}p_{\omega, t}=f_{p}\left(y_{\omega, t}^{i}; \theta_{p}\right), \end{equation}
 for  $i=1,2$ and $\omega  = 1,2, \ldots ,\Omega $.

  We then calculate the average predicted distribution over all the $\Omega$ links in $\Gamma_{lb}\left( t \right)$ as the final predicted distribution:
\begin{equation}
{p_t} = {{\sum\limits_{\omega = 1}^\Omega {{p_{{\omega},t}}} } \mathord{\left/
 {\vphantom {{\sum\limits_{\omega = 1}^Q {{p_{{\omega},t}}} } \Omega}} \right.
 \kern-\nulldelimiterspace} \Omega}.\end{equation}

 \begin{figure*}[!htpb]
	\centering{\includegraphics[width=1\linewidth]{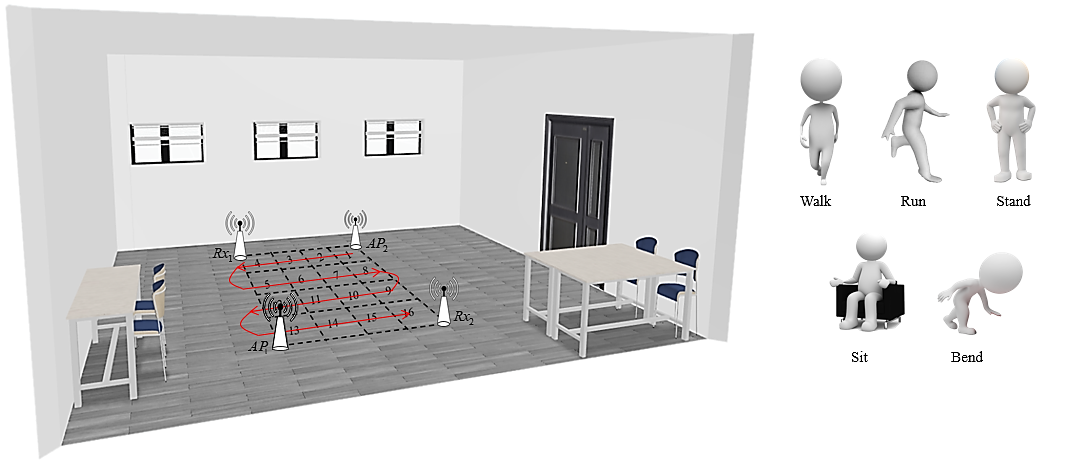}}
	\caption{Illustration of the experimental setting. Left: The $8m \times 7m$ environment, in which the transceivers are located and the 16 locations and the trajectory followed by the target are shown. Right: The five types of activities performed by the target.}
	\label{setup}
\end{figure*}
 The reward will be calculated from the predicted distribution $p_{t}$, taking into account the performance of link selection.
 In this paper, we propose a strategy which is inspired from recent work in the area of computer vision where authors in \cite{wu2019multi} have addressed a problem of video frame selection for lowering the complexity of video processing. Albeit driven by completely independent applications and data scenarios, the problems of frame selection and Wi-Fi link selection bear some interesting analogies in that the objective is to identify a sub-sampled version of an original measured data that capture the greatest amount of relevance to a given data-drive classification problem.
 The goal of the reward function is to push the agent toward finding an effective link group that will gradually improve the classification accuracy.
 In other words, the reward is set to provoke the largest possible prediction gains (in identifying the correct ground-truth class) as compares with the previous iteration \cite{wu2019multi}. Hence, we set the reward received by the agent at the ${t_{th}}$ step as:
 \begin{equation}R_{t}=p_{t,gt}-p_{t-1,gt},\end{equation}
 where $p_{t,\delta}$ is the probability value of class $\delta$ at step ${t_{th}}$, and $gt$ represents the ground-truth label of the behavior.

  Inspired by the REINFORCE \cite{williams1992simple} algorithm of policy gradient, our agent objective is defined as:
  \begin{equation}\label{4}J\left(\theta_{\pi}\right)= \sum_{u \in \mathcal{U}} \pi\left(u | s ;\theta_{\pi}\right) G,\end{equation}
  where $G$ is a long-term discounted reward. Since the link selection is considered as a sequential decision-making problem, a long-term discounted reward is more suitable here, i.e., the further the reward, the lower the contribution to the current step.

 To be specific, the discounted return at the ${t_{th}}$ step can be expressed as:
	 \begin{equation}\label{G} G_{t}=\sum_{k=0}^{t} \psi^{k} R_{t-k},\end{equation}
	 where $\psi$ is the discount rate. As can be seen from Equation (\ref{G}), the parameter gets the most movement in  directions, so that the favorable action can get the highest reward.

	\subsection{Objective Function}
		For one thing, the objective of the agent is to maximize its expected reward.
		For the other thing, we ought to minimize the loss of the classification network.
		We utilize the REINFORCE algorithm to updates the agent's parameters $\theta_{\pi}=\left\{\theta_{o}, \theta_{h}, \theta_{u}\right\}$ and optimize the parameters $\theta_{p}$ with the error Back Propagation (BP) algorithm.

		%

		\subsubsection{Policy Gradient}

		We aim at learning the parameters $\theta_{\pi}$ which can maximize Equation (\ref{4}).
		 The gradient of $J\left(\theta_{\pi}\right)$ is:
		\begin{equation}\label{5}\nabla_{\theta_{\pi}} J\left(\theta_{\pi}\right)=\sum_{u \in \mathcal{U}} \pi\left(u | s ; \theta_{\pi}\right) \nabla_{\theta_{\pi}} \log \pi\left(u | s ; \theta_{\pi}\right) G.\end{equation}
  As the dimension of the action space is high, Equation (\ref{5}) results in a non-trivial optimization problem.
	Therefore, according to REINFORCE, a Monte-Carlo sampling method is used to estimate the gradient:
	\begin{equation}\nabla_{\theta_{\pi}} J\left(\theta_{\pi}\right) \approx \frac{1}{K} \sum_{k=1}^{K}  \sum_{t=0}^{T} \nabla_{\theta_{\pi}} \log \pi\left(u_{t, k} | s_{t, k} ; \theta_{\pi}\right) G_{t},\end{equation}
		where $K$ is the number of samples.
		Via Monte-Carlo stochastic gradient descent, we can minimize the loss function, i.e., Equation (\ref{J}), to updates the parameters $\theta_{\pi}$.
	\begin{equation} \label{J} \mathcal{L}_{J}\left(\theta_{\pi}\right) \approx -\frac{1}{K} \sum_{k=1}^{K}  \sum_{t=0}^{T}\log \pi\left(u_{t, k} | s_{t, k} ; \theta_{\pi}\right) G_{t}.\end{equation}

		\subsubsection{Cross Entropy}
		We define the loss as the mean of binary cross entropy loss for
		each action in the action space:
			\begin{equation}\begin{aligned}
			\mathcal{L}_{U}\left(\theta_{\pi}\right)=&-\frac{1}{M \times Q} \sum_{i=1}^{M \times Q} \sum_{t=0}^{T}\left[\hat{u}_{t, \xi} \log \left(u_{t, \xi}\right)\right.\\
			&\left.+\left(1-\hat{u}_{t, \xi}\right) \log \left(1-u_{t, \xi}\right)\right],
			\end{aligned}\end{equation}
			where ${u}_{t,\xi}$ is the probability of ${\xi_{th}}$ link at the ${t_{th}}$ step, $\hat{u}_{t,\xi}$ is the true value of ${\xi_{th}}$ link at the ${t_{th}}$ step, which can only be 0 or 1.
			Therefore, the loss of the agent is a weighted sum of the two losses:
			\begin{equation}\mathcal{L}_{agent}\left(\theta_{\pi}\right)={\lambda}_{1}\mathcal{L}_{J}\left(\theta_{\pi}\right)
			+ \mathcal{L}_{U}\left(\theta_{\pi}\right),\end{equation}
			where ${\lambda}_{1} $ is a constant.

			\begin{table}[!t]
				\caption{Architecture of the mixed CNN-LSTM network.}
				\label{tA_CNNLSTM}
				\centering
				\fontsize{8}{8}\selectfont
				\setlength{\tabcolsep}{8pt}
				\begin{tabular}{ccccc}
					\toprule
					\textbf{Layer}&\textbf{Name}&\textbf{Size}&\textbf{Strides}&\textbf{Activation} \cr\midrule
					\textbf{1}&Conv 1D&$20 \times 1(8)$&2&ReLU\cr
					\textbf{2}&Conv 1D&$20 \times 1(16)$&2&ReLU\cr
					\textbf{3}&Conv 1D&$20 \times 1(32)$&2&ReLU\cr
					\textbf{4}&Conv 1D&$20 \times 1(64)$&2&ReLU\cr
					\textbf{5}&Max pooling&$2 \times 1$&2&None\cr
					\textbf{6}&LSTM&$128$&None&Tanh\cr
					\textbf{7}&LSTM&$128$&None&Tanh\cr
					\textbf{8}&LSTM&$128$&None&Tanh\cr \bottomrule
				\end{tabular}
			\end{table}
			\begin{table}[!t]
				\caption{Architecture of \emph{CNN 1}.}
				\label{tA_CNN4}
				\centering
				\fontsize{8}{8}\selectfont
				\setlength{\tabcolsep}{8pt}
				\begin{tabular}{ccccc}
					\toprule
					\textbf{Layer}&\textbf{Name}&\textbf{Size}&\textbf{Strides}&\textbf{Activation} \cr\midrule
					\textbf{1}&Conv 2D&$5 \times 5(16)$&2&ReLU\cr
			  \textbf{2}&Dropout& & rate=0.3 & \cr
					\textbf{3}&Max pooling&$2 \times 2$&2&None\cr
			  \textbf{4}&Dropout& & rate=0.5 & \cr
					\textbf{5}&FC&$5$&None&Softmax\cr\bottomrule
				\end{tabular}
			\end{table}
			\begin{table}[!t]
			\caption{Architecture of \emph{CNN 2}.}
			\label{tA_CNN3}
			\centering
			\fontsize{8}{8}\selectfont
			\setlength{\tabcolsep}{8pt}
			\begin{tabular}{ccccc}
				\toprule
				\textbf{Layer}&\textbf{Name}&\textbf{Size}&\textbf{Strides}&\textbf{Activation} \cr\midrule
				\textbf{1}&Conv 2D&$5 \times 5(16)$&2&ReLU\cr
				\textbf{2}&Max pooling&$2 \times 2$&2&None\cr
				\textbf{3}&Conv 2D&$5 \times 5(32)$&2&ReLU\cr
				\textbf{4}&Dropout& & rate=0.3 & \cr
				\textbf{5}&Max pooling&$2 \times 2$&2&None\cr
				\textbf{6}&Dropout& & rate=0.5 & \cr
				\textbf{7}&FC&$5$&None&Softmax\cr\bottomrule
			\end{tabular}
			\end{table}
			\subsubsection{Classification Objective}
			We also define the loss as the mean of binary cross entropy loss for
			each class:
      \begin{equation}
        \mathcal{L}_{P}\left(\theta_{p}\right)=\!-\frac{1}{C}\! \sum_{\delta=1}^{C} \hat{p}_{\delta} \log \left(p_{\delta}\right),
      \end{equation}
			where $C$ is the number of classes, ${p}_{\delta}$ is the probability of ${\delta_{th}}$ class, $\hat{p}_{\delta}$ is the true value of ${\delta_{th}}$ class, which can only be 0 or 1.

	The overall objective of our model is to minimize the loss function:
	\begin{equation}\mathcal{L}oss=\mathcal{L}_{P}\left(\theta_{p}\right)+
	 {\lambda}_{2}\mathcal{L}_{agent}\left(\theta_{\pi}\right),\end{equation}
	where ${\lambda}_{2} $ is a constant.

	\begin{table}[!t]
	\caption{Architecture of \emph{CNN 3}.}
	\label{tA_CNN2}
	\centering
	\fontsize{8}{8}\selectfont
	\setlength{\tabcolsep}{8pt}
	\begin{tabular}{ccccc}
	  \toprule
	  \textbf{Layer}&\textbf{Name}&\textbf{Size}&\textbf{Strides}&\textbf{Activation} \cr\midrule
	  \textbf{1}&Conv 2D&$5 \times 5(16)$&2&ReLU\cr
	  \textbf{2}&Max pooling&$2 \times 2$&2&None\cr
	  \textbf{3}&Conv 2D&$5 \times 5(32)$&2&ReLU\cr
	  \textbf{4}&Dropout& & rate=0.3 & \cr
	  \textbf{5}&Max pooling&$2 \times 2$&2&None\cr
	  \textbf{6}&Conv 2D&$5 \times 5(64)$&2&ReLU\cr
	  \textbf{7}&Dropout& & rate=0.3 & \cr
	  \textbf{8}&Max pooling&$2 \times 2$&2&None\cr
	  \textbf{9}&Dropout& & rate=0.5 & \cr
	  \textbf{10}&FC&$5$&None&Softmax\cr\bottomrule
	\end{tabular}
	\end{table}

	\begin{table}[!t]
		\caption{Architecture of \emph{CNN 4}.}
		\label{tA_CNN1}
		\centering
		\fontsize{8}{8}\selectfont
		\setlength{\tabcolsep}{8pt}
		\begin{tabular}{ccccc}
			\toprule
			\textbf{Layer}&\textbf{Name}&\textbf{Size}&\textbf{Strides}&\textbf{Activation} \cr\midrule
			\textbf{1}&Conv 2D&$5 \times 5(16)$&2&ReLU\cr
			\textbf{2}&Max pooling&$2 \times 2$&2&None\cr
			\textbf{3}&Conv 2D&$5 \times 5(32)$&2&ReLU\cr
	  \textbf{4}&Dropout& & rate=0.3 & \cr
			\textbf{5}&Max pooling&$2 \times 2$&2&None\cr
			\textbf{6}&Conv 2D&$5 \times 5(64)$&2&ReLU\cr
	  \textbf{7}&Dropout& & rate=0.3 & \cr
			\textbf{8}&Max pooling&$2 \times 2$&2&None\cr
			\textbf{9}&Conv 2D&$5 \times 5(64)$&2&ReLU\cr
	  \textbf{10}&Dropout& & rate=0.3 & \cr
			\textbf{11}&Max pooling&$2 \times 2$&2&None\cr
	  \textbf{12}&Dropout& & rate=0.5 & \cr
			\textbf{13}&FC&$400$&None&ReLU\cr
	  \textbf{14}&Dropout& & rate=0.5 & \cr
			\textbf{15}&FC&$5$&None&Softmax\cr\bottomrule
		\end{tabular}
	\end{table}
	\section{Experiments and Implementation}
  To validate the performance of our framework, we develop a proof-of-concept prototype and apply the proposed link selection method to recognize diverse activities, i.e., 5 activities in 16 different locations. We perform a comparison with other possible approaches in the context of commodity Wi-Fi namely random and exhaustive link selection (i.e., all Wi-Fi links are exploited for classification, or possibly a subset of links based on a classical selection method such as selecting links that form an orthogonal subset \cite{qian2017inferring}) various layer architectures in order to understand the impact on effectiveness. Note that our framework can scale to an arbitrary number of deployed APs with reasonable complexity. However due to logistic constraints, we hereby limit our tests to the case of two APs and two Rxs (4 links in total). The details of the experiments and implementation are shown as follows.

\subsection{Experiment Setting}

	In our experiments, we use two mini-PCs as APs and two mini-PCs as Rxs, which are distributed in a square shape as shown in Figure \ref{setup}. The devices are equipped with Intel Wireless Link 5300 NICs and installed CSI-Tool to measure CSI \cite{halperin2011tool}. The Intel 5300 NIC reports 30 out of 56 subcarriers for each of its antennas.
  The APs use one antenna while the Rxs are equipped with three antennas. Note that all the antennas are off-the-shelf horizontally-polarized omni-directional antennas. The transceivers work in the $5.280GHz$ frequency band with $20MHz$ bandwidth. The sampling rate of CSI defaults to $200 Hz$. In order to achieve a multiple legacy Wi-Fi link environment, both two Rxs can receive packets from two APs at the same time, thus forming four Wi-Fi links. In this paper, all APs are not subject to any strict time synchronization. Note that this does not affect the recognition results.

	The environment is about $8m \times 7m$, as shown in Figure \ref{setup}. Other existing Wi-Fi networks such as the campus network are operated as usual during the whole experiments.


	\begin{table}[!t]
		\caption{Parameters of our model.}
		\label{tp_model}
		\centering
		\fontsize{8}{8}\selectfont
		\setlength{\tabcolsep}{30pt}
		\begin{tabular}{cc}
			\toprule
			\textbf{Parameter}&\textbf{Value} \cr\midrule
			${\lambda}_{1}$ &$0.1$\cr
			${\lambda}_{2}$ &$0.1$\cr
			Discount rate($\psi$) & $0.9$ \cr
			Learning rate ($\alpha$) &$0.0001$ \cr
			Batch size ($n$) & $128$ \cr
			Infinitesimal parameter($\epsilon$) &$10^{-8}$ \cr
			\bottomrule
		\end{tabular}
	\end{table}

	\begin{figure}[!t]
	\centering{\includegraphics[width=0.99\linewidth]{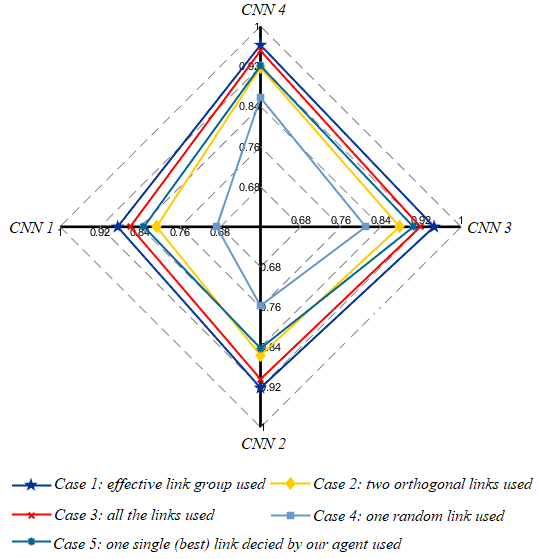}}
	\caption{Radar graph of the overall accuracies for the various algorithm cases and CNN types.}
	\label{radar1}
	\end{figure}
	\subsection{Data Sets}
	As shown in Figure \ref{setup}, we divide the area surrounded by these Wi-Fi links into 16 locations, and the target randomly performs five daily activities of ``Walk", ``Run", ``Stand", ``Sit", ``Bend".
	We collect 2749 real CSI samples in the experiments, with 1949 samples as training and 800 samples as test data. In the test sets, there are 50 samples at each location. It should be noted that the target performs the same activity at all locations across the 16 locations in the trajectory in Figure \ref{setup}, and then performs another run of the trajectory with another activity.

	\subsection{Comparison Experiments}
  We compare the five following strategies for exploiting the various available Wi-Fi links:
	\begin{itemize}
	\item \emph{Case 1: Use the link group $\Gamma_{lb}$ decided by our agent.}
	\item \emph{Case 2: Use two Wi-Fi orthogonal links.}
	\item \emph{Case 3: Use all the Wi-Fi links.}
	\item \emph{Case 4: Use one random Wi-Fi link.}
	\item \emph{Case 5: Use one single (best) link decided by our agent.}
	\end{itemize}

  Hence, the possible benefits of our method is illustrated in \emph{Case 1} and \emph{Case 5}. For all cases, we utilize the CSI of different Wi-Fi links as the data sets for human activity recognition.
	Note that \emph{Case 2} is a classic heuristic approach that utilizes a pair of orthogonal transceivers which has been widely proved to be effective in many state-of-the-art studies \cite{qian2017inferring, wang2017wifall, guo2020from, wang2021from}.
  Furthermore, \emph{Case 4} and \emph{Case 5} use one Wi-Fi link which is a common practice in many current works. Finally note that \emph{Case 4} and \emph{Case 5} exhibit the same complexity.
	\begin{table}[!t]
		\caption{Parameters of \emph{Case 2}, \emph{Case 3} and \emph{Case 4}}
		\label{case2}
		\centering
		\fontsize{8}{8}\selectfont
		\setlength{\tabcolsep}{30pt}
		\begin{tabular}{cc}
			\toprule
			\textbf{Parameter}&\textbf{Value} \cr\midrule
			Learning rate ($\alpha$) &$0.001$ \cr
			Batch size ($n$) & $128$ \cr
			Maximum epoch ($e$) & $200$ \cr
			Infinitesimal parameter($\epsilon$) &$10^{-8}$ \cr
			\bottomrule
		\end{tabular}
	\end{table}

	\begin{figure}[!t]
	\centering{\includegraphics[width=0.999\linewidth]{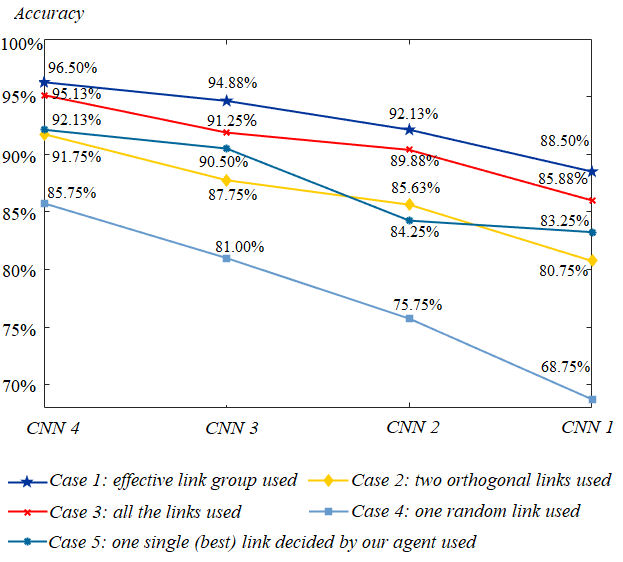}}
	\caption{Accuracy versus the number of CNN-based classification network layers.}
	\label{line1}
	\end{figure}

	\subsection{Network Implementation}
  The network is implemented with TensorFlow \cite{abadi2016tensorflow} and the workstation for our experiments is equipped with 48 core CPUs of Intel Xeon(R) $2.50GHz$, 4GPUs of NVIDIA GeForce GTX1080Ti and 48GB memory. 

	\begin{figure*}[htpb]
		\centering{\includegraphics[width=0.98\linewidth]{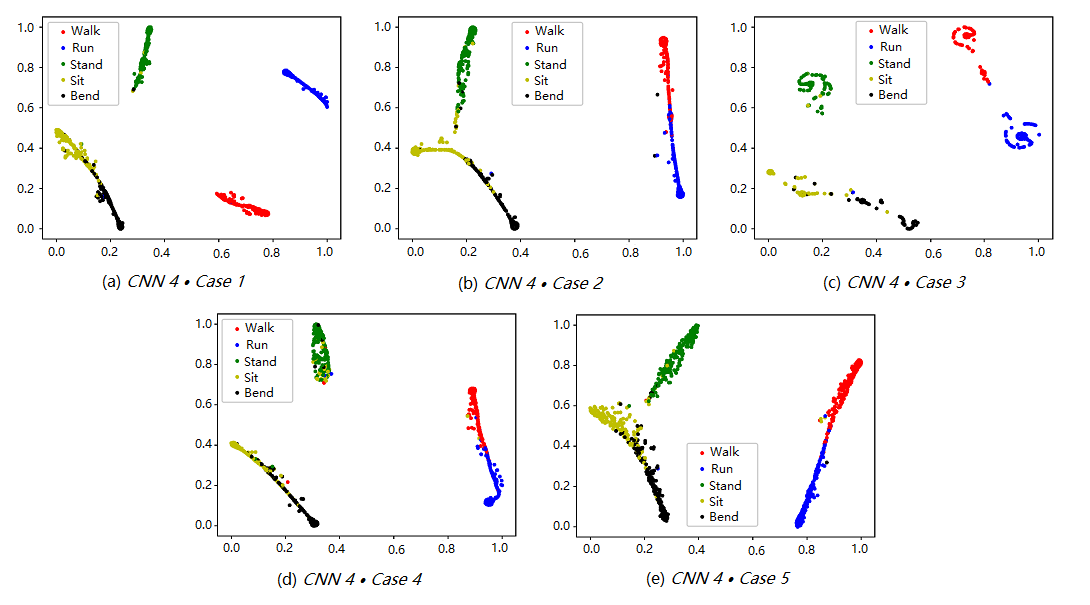}}
	  \caption{The t-SNE embeddings transformed by the last layer features of \emph{CNN 4} in the five different cases. (a) to (e) represent the t-SNE embedding outputs of Case 1 to Case 5, respectively. As can be seen, our framework (Case 1) gives the best distinction between the activity types in the embeddings.}
	  \label{draw_tsne}
	\end{figure*}
%
	For \emph{Case 1} and \emph{Case 5}, as the input data of the agent, the normalized CSI streams are processed in the mixed CNN-LSTM network.
	TABLE \ref{tA_CNNLSTM} shows the architecture of the network. We note that \emph{Case 5} uses a softmax activation function to output the probability value of each action instead of the sigmoid function in the policy network.
	The architectures of the CNN-based classification networks with various layers are presented in TABLE \ref{tA_CNN4} (\emph{CNN 1}), TABLE \ref{tA_CNN3}  (\emph{CNN 2}), TABLE \ref{tA_CNN2}  (\emph{CNN 3}) and TABLE \ref{tA_CNN1}  (\emph{CNN 4}).
	Other parameters of our model are shown in TABLE \ref{tp_model}.


  \emph{Case 2}, \emph{Case 3} and \emph{Case 4} utilize networks with the same architectures but different parameters as the CNN-based classification networks for \emph{Case 1} and \emph{Case 5}. TABLE \ref{case2} shows the parameters of \emph{Case 2}, \emph{Case 3}, and \emph{Case 4}. Note that below each of the experimental settings are numbered in the following way: \emph{CNN c $\bullet$ Case a}, where \emph{Case a} refers to one of the methods shown in Section V. C. with $a = 1; 2; 3; 4; 5$, and \emph{CNN c} refers to the \({c_{th}}\) CNN architecture, $c = 1; 2; 3; 4$.

\section{Performance Evaluation}
\subsection{Evaluation Metrics}
We evaluate the performance of our link selection framework based on the following metrics: 
\textbf{(i)} Overall Effectiveness.
\textbf{(ii)} Characteristic Distribution. \textbf{(iii)} Confusion Matrix of Accuracy. \textbf{(iv)} Computational Cost. We also test those metrics against various scenarios, including changing the number of selected links or varying the target location.



\subsection{Experimental Results}
\subsubsection{Overall Effectiveness} We explore the relationship between different types of CNN-based classification networks and the overall accuracies of human activity recognition in the five cases. The overall classification accuracies of \emph{Case a} \((a = 1;2;3;4;5)\) under \emph{CNN c} \((c = 1;2;3;4)\) are expressed through a radar graph (Figure \ref{radar1}) firstly. Through the radar graph,
we can see that the performance of \emph{Case 1} is the best, and the performance of \emph{Case 5} mostly exceeds that of \emph{Case 2} even if it uses less CSI data.


Then we draw the accuracy curves of different cases with a decreasing number of CNN network layers, as shown in Figure \ref{line1}. It can be seen that the accuracy of \emph{Case 1} declines the slowest and remains at the highest level from Figure \ref{line1}. We can also see that the accuracy of \emph{CNN 2 $\bullet$ Case 1} is much higher than that of \emph{CNN 3 $\bullet$ Case 3}, the accuracy of \emph{CNN 3 $\bullet$ Case 1} is comparable to that of \emph{CNN 4 $\bullet$ Case 3}, which shows that the link selection framework can utilize a simpler network to achieve higher accuracy.

Our framework can significantly outperform the baseline methods. Even if only one single link is decided (\emph{Case 5}), the accuracy is higher than using two orthogonal links (\emph{Case 2}). Furthermore, the numbers of Wi-Fi links selected by our agent under \emph{CNN c} \((c = 1;2;3;4)\) are 2.90, 3.14, 2.49 and 2.35, respectively.

Due to the limited space, in what follows below we only focus on \emph{CNN 4},
as it seems to give the best compromise between complexity and performance.

\begin{figure*}[htpb]
	\centering{\includegraphics[width=0.99\linewidth]{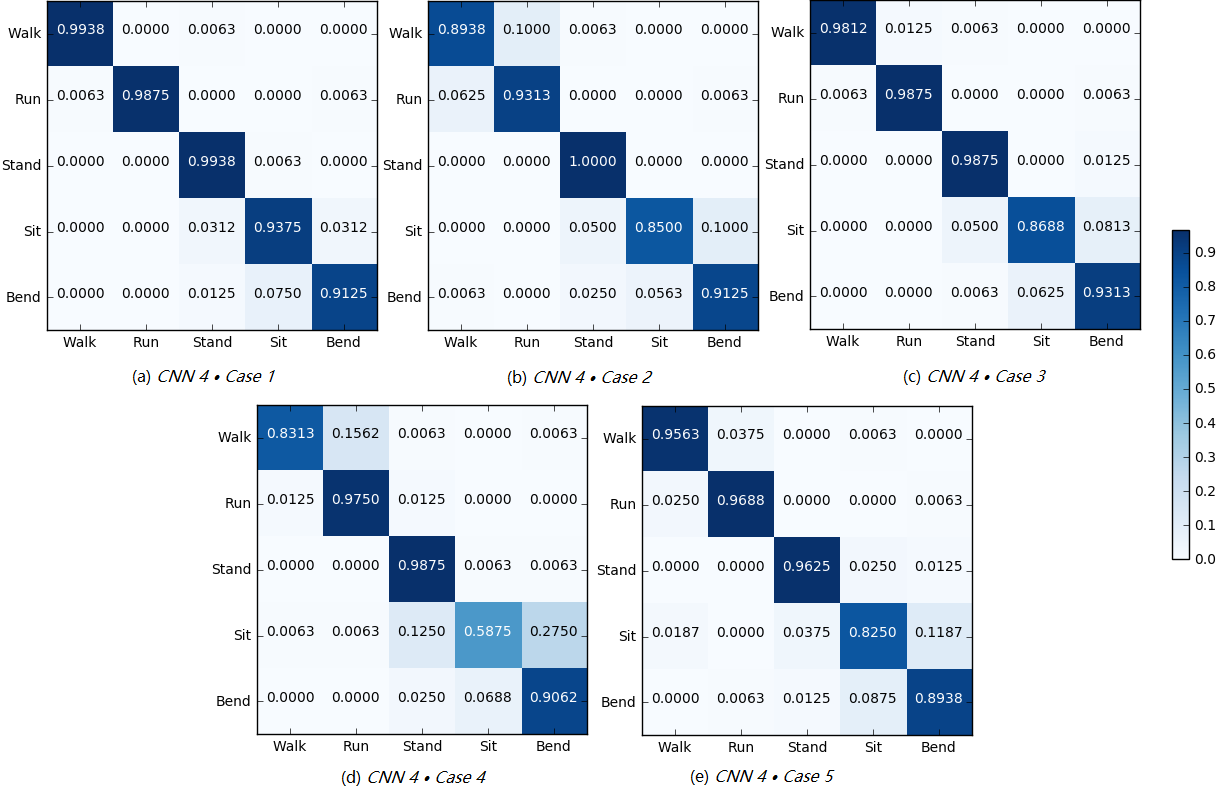}}
	\caption{Confusion matrix of the Wi-Fi based human activity recognition accuracy in different cases under \emph{CNN 4}. (a) to (e) represent the confusion matrices of Case 1 to Case 5, respectively. As can be seen, our framework (Case 1) achieves the best performance.}
	\label{draw_CM}
	\end{figure*}
\subsubsection{Characteristic Distribution}
We examine the characteristics obtained by our proposed framework that underpins the successful performance of the agent in the Wi-Fi-based human activity recognition.
The t-SNE algorithm is a technique developed for the visualization of high-dimensional data \cite{mnih2015human}. As expected, the t-SNE embeddings tend to map the characteristics of perceptually similar activities to nearby points.
Thus we can utilize the t-SNE embeddings to estimate the performance of the human activity recognition.
As shown in Figure \ref{draw_tsne}, we plot the t-SNE embeddings transformed by the last layer features of the CNN classification network (\emph{CNN 4}) in different cases.

The higher performance achieved by our method (\emph{Case 1}) appears clearly from Figure \ref{draw_tsne}(a). In that plot, the embeddings associated with each one of the 5 activities appear as fairly compact and well distinct groups of points, hence helping the classification task.

In contrast, the other methods result in embeddings that show up as groups of points with significant overlap or tend to be more scattered. This effect is particularly pronounced in \emph{Case 4}, shown in Figure \ref{draw_tsne}(d), where it is seen that totally random selection of one of the Wi-Fi links gives the worst performance. In comparison the selection of just one link but based on our RL strategy (\emph{Case 5}) gives a clear improvement over \emph{Case 4}, as see in Figure \ref{draw_tsne}(e). Interestingly, \emph{Case 3}, which systematically exploits all radio links performs better that a single random random link approach but not as well as a carefully selected subset of links, as shown in Figure \ref{draw_tsne}(c) because some of the links tend to be more noisy or carry data less relevant to the classification task at hand, hence tend to confuse the network. Finally \emph{Case 2} corresponds to a widely used approach to heuristically select two radio links (based on orthogonality \cite{qian2017inferring, guo2020from, wang2021from}), but again results in greater confusion probability as shown in Figure \ref{draw_tsne}(b) when compared to our \emph{Case 1} in Figure \ref{draw_tsne}(a).

	\begin{figure*}[htpb]
	  \centering{\includegraphics[width=0.99\linewidth]{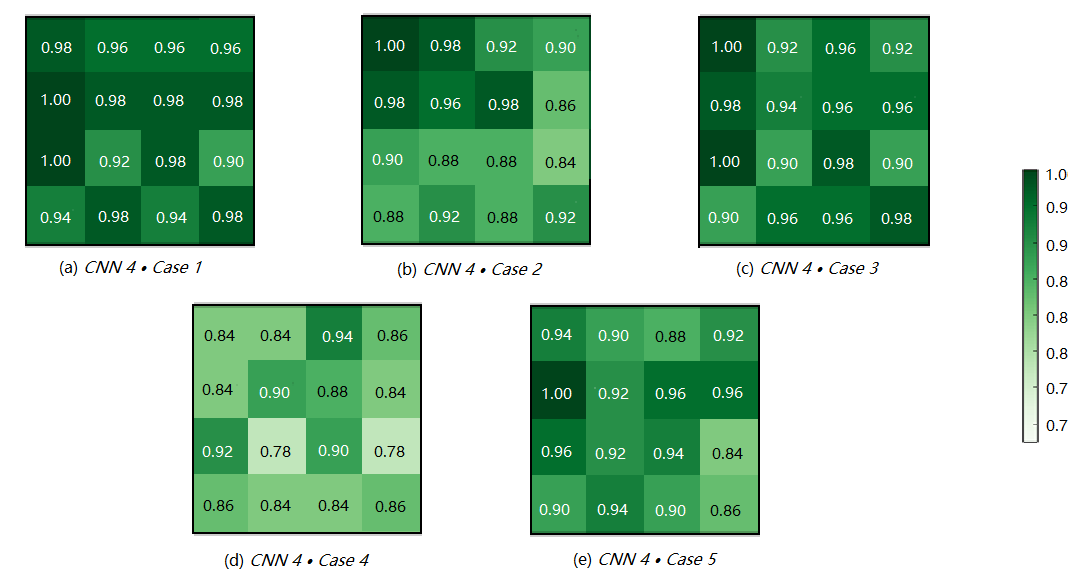}}
	  \caption{Accuracy of the Wi-Fi based human activity recognition at different locations in different cases under \emph{CNN 4}. (a) to (e) represent the accuracy at different locations of Case 1 to Case 5, respectively. As can be seen, our framework (Case 1) achieves the best performance.}
	  \label{draw_LOC}
	\end{figure*}
\subsubsection{Confusion Matrix of Accuracy}
To evaluate a fine-grained performance, we present the confusion matrix of accuracy for each case under \emph{CNN 4} in Figure \ref{draw_CM}.

\emph{Case 1:} Figure \ref{draw_CM}(a) shows the confusion matrix of accuracy for \emph{CNN 4 $\bullet$ Case 1}.
Obviously, our proposed framework outperforms all others, reaching a recognition rate of 96.50$\%$.
To be specific, the accuracy of each activity is 99.38$\%$, 98.75$\%$, 99.38$\%$, 93.75$\%$ and 91.25$\%$, respectively.
The accuracy of each activity ranges from 91$\%$ to 100$\%$, indicating that the effective link group $\Gamma_{lb}$ decided by our agent can not only accurately but also comprehensively classify the overwhelming majority of corresponding activity.

\emph{Case 2:} Figure \ref{draw_CM}(b) shows the confusion matrix of accuracy for \emph{CNN 4 $\bullet$ Case 2}.
The overall recognition accuracy for all activities in \emph{Case 2} reaches 91.75$\%$.
Although Case 2 has a similar recognition accuracy to \emph{Case 1} in most activities, it has a poor recognition effect for ``Walk" and ``Sit".
As reflected in existing papers \cite{qian2017inferring, wang2017wifall, guo2020from}, \emph{Case 2} is applicable to some daily activities.
However, it does not guarantee that the two orthogonal Wi-Fi links closest to the human location can extract the optimal activity characteristics in a legacy network of multiple APs.
Especially in the application of wireless signals for human sensing, it is more important to choose links flexibly than to use fixed links.

\emph{Case 3:} Figure \ref{draw_CM}(c) shows the confusion matrix of accuracy for \emph{CNN 4 $\bullet$ Case 3}. The overall recognition accuracy is 95.13$\%$, which is lower than that of \emph{Case 1}.
It can be seen that, ``Sit" samples are more likely to be mistaken for ``Bend", which may be due to the fact that the duration of the two activities are quite similar.
Some Wi-Fi links may not be sensitive enough to small differences in these characteristics, so such links need to be eliminated.

\emph{Case 4:} Figure \ref{draw_CM}(d) shows the confusion matrix of accuracy for \emph{CNN 4 $\bullet$ Case 4}.
We can see that randomly selecting a link is the least effective way to classify activities in all cases, with the overall accuracy of 85.75$\%$.

In conclusion, it is not feasible to use too much or too little information without filtering.

\emph{Case 5:} Figure \ref{draw_CM}(e) shows the confusion matrix of accuracy for \emph{CNN 4 $\bullet$ Case 5}. This case performs slightly worse than \emph{Case 1} and \emph{Case 3}, but better than the other two cases, with the overall recognition accuracy of 92.13$\%$. This illustrates, on the one hand, the limitation of using a single link for human sensing and, on the other hand, the effectiveness of our framework.

%
%

\begin{table*}[htbp]
	\centering
	\caption{Test Time of the five cases using different CNNs}
	\fontsize{8}{8}\selectfont
	\setlength{\tabcolsep}{5.9pt}
		\begin{tabular}{c|cccccccccc}
		\toprule
			\diagbox{\textbf{CNN Type}}{\textbf{Time Cost}}{\textbf{Algorithm Case}}   & \multicolumn{2}{c}{\textit{\textbf{Case 1}}} & \multicolumn{2}{c}{\textit{\textbf{Case 2}}} & \multicolumn{2}{c}{\textit{\textbf{Case 3}}} & \multicolumn{2}{c}{\textit{\textbf{Case 4}}} & \multicolumn{2}{c}{\textit{\textbf{Case 5}}} \\
		\midrule
			& \textbf{Decision} & \multicolumn{1}{l}{\textbf{Classification}} & \multicolumn{2}{c}{\textbf{Classification}} & \multicolumn{2}{c}{\textbf{Classification}} & \multicolumn{2}{c}{\textbf{Classification}} & \textbf{Decision} & \multicolumn{1}{l}{\textbf{Classification}} \\
		\textit{\textbf{CNN 1}} & 15.3ms & 5.3ms & \multicolumn{2}{c}{3.8ms} & \multicolumn{2}{c}{6.3ms} & \multicolumn{2}{c}{2.3ms} & 13.8ms & 2.5ms \\
		\textit{\textbf{CNN 2}} & 15.9ms & 5.5ms & \multicolumn{2}{c}{4.1ms} & \multicolumn{2}{c}{6.3ms} & \multicolumn{2}{c}{2.5ms} & 14.5ms & 2.8ms \\
		\textit{\textbf{CNN 3}} & 14.6ms & 5.1ms & \multicolumn{2}{c}{4.5ms} & \multicolumn{2}{c}{6.6ms} & \multicolumn{2}{c}{2.6ms} & 14.4ms & 2.8ms \\
		\textit{\textbf{CNN 4}} & 15.9ms & 5.2ms & \multicolumn{2}{c}{4.6ms} & \multicolumn{2}{c}{7.3ms} & \multicolumn{2}{c}{2.9ms} & 14.2ms & 2.9ms \\
		\bottomrule
		\end{tabular}%
	\label{Cost}%
	\end{table*}%
	\subsubsection{Recognition Accuracy of Different Locations}
		Then we calculate the human activity recognition accuracy of different locations under the five cases, as shown in Figure \ref{draw_LOC}.

		\emph{Case 1:} Figure \ref{draw_LOC}(a) shows the recognition accuracy of the 16 locations under \emph{CNN 4 $\bullet$ Case 1}. The recognition accuracy of each location reaches more than 90\%, among which the lowest accuracies are in Location 9, but it could still reach 90\%. In contrast, in other cases (except \emph{Case 3}), the recognition accuracy in some locations are less than 90\%, or even less than 80\%. We also note that the performance of \emph{Case 1} is better than that of \emph{Case 3}, as the accuracies of \emph{Case 1} over all locations distribute uniformly and the accuracy of each location is generally higher than that of \emph{Case 3}. This indicates that our framework can select the effective link group ($\Gamma_{lb}$) for different locations ($l$), thus effectively improving the recognition accuracy.

		\emph{Case 2:} Figure \ref{draw_LOC}(b) shows the recognition accuracy of the 16 locations under \emph{CNN 4 $\bullet$ Case 2}. In this case, the recognition accuracy of each location is lower than \emph{Case 1}, especially the Location 8, Location 9, Location 10, Location 11, Location 13, Location 15, whose accuracies are lower than 90\%. This shows that the two orthogonal links are not enough for each location to get the optimal sensing results. In addition, it shows that our framework can effectively improve the recognition accuracy of the difficult areas.

		\emph{Case 3:} Figure \ref{draw_LOC}(c) shows the recognition accuracy of the 16 locations under \emph{CNN 4 $\bullet$ Case 3}. In this case, the recognition accuracies of most locations are lower than \emph{Case 1}. Among them, the locations with the worst recognition accuracies are Location 9, Location 11 and Location 13, all are 90\%. This indicates that there are some redundant links that affect the recognition accuracy, i.e., the links themselves do not contain activity-related features.

		\emph{Case 4:} Figure \ref{draw_LOC}(d) shows the recognition accuracy of the 16 locations under \emph{CNN 4 $\bullet$ Case 4}. This case has the worst performance, which indicates that selecting a single link randomly has obvious disadvantages.

	  \emph{Case 5:} Figure \ref{draw_LOC}(e) shows the recognition accuracy of the 16 locations under \emph{CNN 4 $\bullet$ Case 5}. The performance of this case is slightly worse than \emph{Case 1} and \emph{Case 3}, but better than \emph{Case 2}, illustrating the necessity of link selection. In addition, for some locations, such as Location 2, Location 9 and Location 16, the recognition accuracy is lower, indicating that using one single link for recognition is not enough for these locations, i.e., utilizing only one single link for human sensing has limitations.
		\begin{table}[!t]
			\centering
			\caption{Classification Accuracy vs. Number of Selected Links}
			\centering
					\fontsize{8}{8}\selectfont
					\setlength{\tabcolsep}{2.5pt}
			\begin{tabular}{c|cccccccc}
			\toprule
			\diagbox{\textbf{Number of Links}}{\textbf{CNN Type}} & \multicolumn{2}{c}{\textbf{\emph{CNN 1}}} & \multicolumn{2}{c}{\textbf{\emph{CNN 2}}} & \multicolumn{2}{c}{\textbf{\emph{CNN 3}}} & \multicolumn{2}{c}{\textbf{\emph{CNN 4}}} \\
			\midrule
			\textbf{Best 2 Links Selected}     & \multicolumn{2}{c}{89.13\%} & \multicolumn{2}{c}{90.75\%} & \multicolumn{2}{c}{94.50\%} & \multicolumn{2}{c}{96.25\%} \\
			\textbf{Best 3 Links Selected}      & \multicolumn{2}{c}{89.88\%} & \multicolumn{2}{c}{91.75\%} & \multicolumn{2}{c}{95.38\%} & \multicolumn{2}{c}{96.88\%} \\
			\textbf{Undetermined Links}  & \multicolumn{2}{c}{88.50\%} & \multicolumn{2}{c}{92.13\%} & \multicolumn{2}{c}{94.88\%} & \multicolumn{2}{c}{96.50\%} \\
			\textbf{All Links} & \multicolumn{2}{c}{85.88\%} & \multicolumn{2}{c}{89.88\%} & \multicolumn{2}{c}{91.25\%} & \multicolumn{2}{c}{95.13\%} \\
			\bottomrule
			\end{tabular}%
			\label{Acc23}%
		\end{table}%

	  \subsubsection{Classification Accuracy vs. Number of Selected Links}

	  To further evaluate the effectiveness of our framework, we make the agent select a fixed number of links in $\Gamma_{lb}$ and then calculate the recognition accuracy. That is, our agent decides the best two or three links for classification each time. The experiment results under different layers of CNN-based classification networks are shown in TABLE \ref{Acc23}. The results demonstrate the effectiveness of our framework.


\subsubsection{Computational Cost}

Finally, we evaluate the computational cost of the five cases using different layers of CNN-based classification networks. We calculate the average test time over all the test samples and the results are shown in TABLE \ref{Cost}.

For one thing, the link selection process takes about 15$ms$, which is still in the acceptable range. Further, our framework brings a high gain in accuracy, which illustrates the significance of our framework.

For the other thing, the time required for the link selection process is constant. Because of the fixed numbers of links used, the test time of the classified network in \emph{Case 2}-\emph{Case 4} increases with the complexity of the network. Therefore, in more densely deployed Wi-Fi environments, our framework will be more advantageous.

	\section{Discussion and Future work}
  In this paper, we present a RL-based link selection framework that is capable of identifying the most relevant group of radio links from multiple APs for human behavior sensing.

	Our experimental results show that our proposed framework can achieve 96.50\% accuracy when utilized to recognize 5 daily activities at 16 different locations in a Wi-Fi environment with up to four radio links available. This accuracy level is higher than what is obtained by using the data from all radio links, and is much higher than what is obtained with an arbitrary or random selection of links. In the given environment, our proposed method only relies on 2.35 links per sample on average as shown in Table \ref{Num-link}. This table also indicates the average number of radio links exploited per different types of activity and human locations. As can be seen, some dynamic activities such as ``Walk" and ``Run" tend to rely on a higher number of links while more static ones can live with a fewer number of links and data. Our current setup is limited to four radio links but is scalable in practice. We believe performance will further increase with a higher number of links expected from a real-life dense Wi-Fi deployment.

	\begin{table}[!t]
	\centering
	\caption{The average numbers of links in the $\Gamma_{lb}$ selected by the RL agent, depending on human location and activity type.}
	\fontsize{8}{8}\selectfont
	\setlength{\tabcolsep}{5.1pt}
	  \begin{tabular}{ccccccc}
	  \toprule
	  \textbf{Human Location} & \textbf{Walk} & \textbf{Run} & \textbf{Stand} & \textbf{Sit} & \textbf{Bend} & \textbf{Average} \\
	  \midrule
	  \textbf{1} & 2.70  & 3.10  & 2.20  & 2.00  & 2.20  & 2.44  \\
	  \textbf{2} & 2.90  & 2.90  & 1.80  & 2.10  & 2.10  & 2.36  \\
	  \textbf{3} & 2.30  & 3.30  & 2.30  & 1.90  & 1.80  & 2.32  \\
	  \textbf{4} & 2.30  & 2.90  & 2.50  & 2.40  & 2.00  & 2.42  \\
	  \textbf{5} & 2.20  & 2.60  & 2.30  & 2.50  & 2.20  & 2.36  \\
	  \textbf{6} & 2.70  & 2.90  & 1.80  & 2.10  & 2.00  & 2.30  \\
	  \textbf{7} & 2.30  & 2.40  & 2.30  & 2.50  & 2.40  & 2.38  \\
	  \textbf{8} & 2.60  & 2.20  & 2.30  & 2.40  & 2.20  & 2.34  \\
	  \textbf{9} & 2.50  & 1.90  & 1.60  & 2.00  & 2.30  & 2.06  \\
	  \textbf{10} & 2.60  & 2.80  & 2.00  & 2.10  & 2.40  & 2.38  \\
	  \textbf{11} & 2.70  & 2.20  & 2.20  & 2.20  & 2.40  & 2.34  \\
	  \textbf{12} & 2.40  & 3.00  & 1.90  & 2.60  & 2.40  & 2.46  \\
	  \textbf{13} & 2.80  & 2.60  & 2.00  & 1.90  & 2.60  & 2.38  \\
	  \textbf{14} & 2.80  & 2.10  & 1.80  & 1.90  & 2.10  & 2.14  \\
	  \textbf{15} & 2.50  & 3.10  & 2.00  & 2.40  & 2.20  & 2.44  \\
	  \textbf{16} & 2.50  & 2.90  & 2.20  & 2.10  & 2.40  & 2.42  \\
  \cmidrule{1-1}    \textbf{Average} & 2.55  & 2.68  & 2.08  & 2.19  & 2.23  & 2.35  \\
	  \bottomrule
	  \end{tabular}%
	\label{Num-link}%
  \end{table}%
	Based on our proposed framework, one can continue to improve its functionality to deal with more diverse application scenarios, such as human imaging, or more complex Wi-Fi environments including through-wall setups.
\section*{Acknowledgment}
This work was supported by Beijing Nova Program from Beijing Municipal Science \& Technology Commission under Grant Z201100006820123 and National Key R \& D Program of China under Grant 2018YFC0810204. The authors thank the volunteers for participating in the experiments. Finally, the authors sincerely thank the anonymous reviewers for their insightful comments.

\ifCLASSOPTIONcaptionsoff
  \newpage
\fi

\bibliographystyle{IEEEtran}
\bibliography{IEEEabrv,Ref-new}

\end{document}